# A hidden anti-jamming method based on deep reinforcement learning

Yifan Wang[1], Xin Liu[1,2], Mei Wang[1] and Yu Yu[3]

[1] Guilin University of Technology College of Information Science and Engineering Guilin, China

[2] Guilin University of Technology Guangxi key Laboratory Fund of Embedded Technology and Intelligent System Guilin, China

[3] 31131 Troops, PLA, China

[e-mail: liuxin2017125@glut.edu.cn]

*Corresponding author: Xin Liu

This research work was supported by the National Natural Science Foundation of China under Grant 61961010, the Key Laboratory Found of Cognitive Radio and Information Processing, Ministry of Education (Guilin University of Electronic Technology) under Grant No. CRKL180203，the 'Ba Gui Scholars' program of the provincial government of Guangxi".

**Abstract**

Most of the current anti-jamming algorithms for wireless communications only consider how to avoid jamming attacks, but ignore that the communication waveform or frequency action may be obtained by the jammers. Although existing anti-jamming methods can guarantee temporary communication effects, the long-term performance of these anti-jamming methods may be depressed when intelligent jammers are capable of learning from historical communication activities. Aiming at this issue, a hidden anti-jamming method based on the idea of reducing the jammer's sense probability is proposed. Firstly, the sensing probability of the jammer is obtained by calculating the correlation between the actions of the jammer and the user. Later, a deep reinforcement learning framework is designed, which aims at not only maximizing the communication throughput but also minimizing the action's correlation between the jammer and the user. Finally, a hidden anti-jamming algorithm is proposed, which links the instantaneous return with the communication quality of users and the correlation between users and jammer. The simulation result shows that the proposed algorithm not only avoids being sensed by the jammer but also improves its anti-jamming performance compared to the current algorithm that only considers jamming avoidance.

# 1. Introduction

Anti-jamming is a hot topic which aims at realizing continuous and stable communications. Although the mobility and openness of wireless communications bring many conveniences to people's lives, there are also a lot of security issues, such as being susceptible to various types of jamming attacks [1]-[3]. In recent years, with the development of artificial intelligence, jamming with environmental sense and learning capabilities have posed new challenges to anti-jamming technology. The anti-jamming technology urgently needs to be iteratively upgraded to more intelligent [4]-[6].

To confront intelligent jammer with environmental sense and learning capabilities, genetic algorithms [7], particle swarm algorithms [8], and artificial bee colony algorithms [9] have been used for making frequency, power, or coding action. However, such algorithms need different degrees of prior information, which limits their application in real scenarios. Considering the interaction between users and malicious jammer, game theory applies to the analysis of communication strategies in the case of jamming and has been used extensively in the field of anti-jamming. For example, spectrum resource allocation and dynamic spectrum access, etc[10]-[14]. Generally, the purpose of the user of the communication is to avoid jamming, while the jammer wants its frequency to be the same as the user. The utility of both parties is exactly the opposite. The "zero-sum game" is often used as a model for frequency immunity[15]. Considering the characteristics of the user and the jammer hierarchical decision making, frequency domain anti-jamming can also be modeled using a Steinberg game[16]. However, decision algorithms based on game-theoretic models rely on many assumptions that are difficult to implement in practice and affect the practical application. To address the reality that the state of the environment is unknown and difficult to get, the anti-jamming technique based on reinforcement learning has been more widely used and achieved more satisfactory results. Examples include adaptive frequency hopping actions based on Q-learning [17], joint time-frequency anti-jamming communication [18], and intelligent anti-jamming relay systems based on reinforcement learning [19]. With the improvement of jamming learning ability, users need to face an increasing amount of state space. Anti-jamming communication based on reinforcement learning is difficult to converge quickly, which affects the quality of communication. A sequential deep reinforcement learning method that uses deep learning to classify complex environmental states, enabling subsequent use of reinforcement learning for optimal action making, is proposed in Liu et al [20]. Although it shortens the convergence time, it can only cope with the dynamic jamming mode, and the anti-jamming ability effect will be weakened as the jammer adopts intelligent jamming strategies.

To counteract intelligent jammer, anti-jamming methods based on deep reinforcement learning have become a popular research direction today. For example, it is proposed in the literature [21]-[23] to use the time-frequency 2D information as the original input and apply the deep reinforcement learning technique against intelligent jammer. However, it makes optimal actions by considering only how to maximize jamming avoidance, i.e., just the SINR of the user is used as a basis for judging the instantaneous returns of the user. DRL is also used in the literature [24] to select the optimal policy. What's different from other articles is it uses a deep deterministic policy gradient (DDPG) instead of stochastic gradient descent (SGD) to update the network parameters. Although it shortens the convergence time, it only relates the instantaneous returns to the user's SINR when making optimal actions, i.e., it only considers whether the user has avoided jamming. In order to reduce the energy consumption of the system, [25] defines the environmental state as 3-dimensional information (time, frequency, and power) and also uses deep reinforcement learning algorithms for optimal action making (frequency and power). The system can guarantee less energy consumption with increased throughput. But the core idea is still how to avoid jamming to the maximum extent possible.

In summary, the core idea of the above methods is to maximize the probability of avoiding jamming. Although a good anti-jamming effect can be achieved at the current moment, the user's past signal waveform and frequency action information may be exposed. As the jamming continues to learn the above information, the effectiveness of anti-jamming may diminish. Therefore, how to cope with intelligent jamming while ensuring that the information of the user will not leak as far as possible is also a direction that anti-jamming researchers need to explore in-depth. Intelligent jamming relies on the sense of the environment for learning actions, and the jammer sense of the environment is influenced by environmental factors. If the user tries to select frequency points where the jammer does not sense well, it may avoid jamming as well as the information leakage. For example, there is a high power station near the jammer but far from the user's receiver, the transmitting frequency of this station is an excellent frequency for users to avoid sensing. Unfortunately, the above information is not available to the user in advance. However, the user can indirectly verify whether the user is avoiding being sensed by the jammer by analyzing the correlation between the jammer's action and the user's action. Because the previous jammer's action and the user's action are known to the user. Therefore, an action relevance method is designed to measure the correlation between the jammer's action and the user's action and to judge whether the user's action can avoid being sensed by the jammer. The action correlation measurement and deep reinforcement learning are combined to realize hidden anti-jamming communications. Finally, an anti-jamming deep reinforcement learning algorithm based on hiding strategy(ADRLH) is proposed.

The main contributions are summarized as follows:

•A hidden anti-jamming idea is proposed. The user uses rational communication action-making to prevent the jammer from obtaining the information of the user so that the jammer cannot target the user for jamming.

•A measurement method for evaluating hide performance is designed in this paper. The evaluation of the hiding effect is difficult owing to the jammer does not actively inform the user of its sensing results. Aiming at this problem, this paper analyzes the hiding effect of users from the perspective of action correlation between the user and the jammer. Obviously, if the jammer can effectively sense the user's actions, then its jamming actions should be highly related to the user's actions, i.e., the reactive jamming can be modeled as a delayed function of the user's actions. Otherwise, if their actions are not relevant, it could be inferred that the jammer doesn't sense the action of the user effectively. Therefore, the correlation between the actions of the user and the jammer can evaluate the user's hidden effect.

•An anti-jamming learning approach aiming at hiding is proposed, which can not only avoid jamming but also reduce information-exposure. Specifically, in addition to the throughput performance, the hidden performance is taken as a part of the instantaneous return, so that the user will prefer to the decision that can conceal its signal after learning. Therefore, even the highly intelligent jammer is not able to make effective jamming strategies due to the lack of valuable sensing information.

The rest part of this paper is presented as follows. In Section 2, the anti-jamming system model is given. After that, an anti-jamming deep reinforcement learning algorithm based on hiding strategy is presented in Section 3. Besides, the analysis of simulation results and the conclusion are given in Section 4 and Section 5 respectively.

## 2. Anti-jamming system model

The anti-jamming system model is schematically illustrated in Fig. 1. It is mainly composed of a transmitter, a receiver, a jammer, a sensor, and an environmental interference source. The transmitter sends the signals to the receiver. The receiver accepts signals from the environment and converts them into environmental information. It learns to make actions and feeds back to the transmitter whether or not to communicate. And it transmits communication frequencies through the next time slot. The sensor passes the signals obtained from the environment to the jammer, and then the jammer makes action and releases the jamming signal at the corresponding frequency point after learning from the above environmental signals.

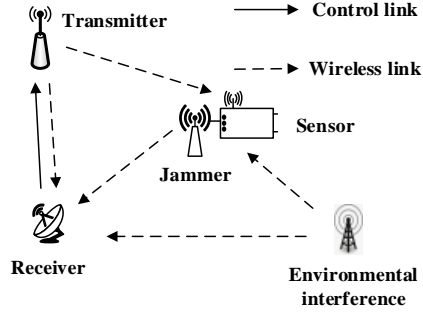

**Fig. 1**. Anti-jamming system model

The starting frequency and ending frequency of the communication band of both user and jammer is $f_s$ and $f_e$ respectively, the number of channels is $N$, and the signal bandwidth of both antagonists is $b = (f_e - f_s)/N$. The center frequency of the signal for the user is selected as $f_t^I \in \{(i-1/2)(f_e - f_s)/N\}$, where $i = \{1, 2, \cdots, N\}$. The transmitting power of the signal is $P^I = \int_{-b/2}^{b/2} U(f) df$, where $U(f)$ is the power spectral density (PSD). The center frequency of the signal for the jammer is chosen as $f_t^J \in \{(i-1/2)(f_e - f_s)/N\}$, and waveform denoted by baseband PSD function is $J(f)$. The PSD of the environment interference signal is $E(f)$, and of the noise signal is $n(f)$. The channel gain of the transmitter-receiver link is represented by $g^{TR}$, and the channel gain of the environmental interference to the receiver is $g^{ER}$. The received SINR of the user can be expressed as:

$$\Phi_t^I = \frac{g^{TR} P^I}{\int_{f_t^I - b/2}^{f_t^I + b/2} \left\{ n(f) + g^{JR} J(f - f_t^J) + g^{ER} E(f - f_t^E) \right\} df} \quad (1)$$

The jammer confirmed the user signal by capturing the synchronization sequence of the user. The lower the user's SINR sensed by the jammer, the lower the accuracy of determining the existence of the user signal. $g^{TJ}$ indicates the channel gain from the transmitter to jammer, and $g^{EJ}$ indicates the channel gain from environmental interference to the jammer. The jammer sense user's SINR can be represented as:

$$\Phi_{t,n}^{J} = \frac{g^{TJ} P^{I}}{\int_{(n-1)b}^{nb} \left\{ \left\{ n(f) + g^{JR} J\left(f - f_{t}^{J}\right) + g^{EJ} E\left(f - f_{t}^{E}\right) \right\} df \right\}} \quad (2)$$

where $n$ represents the number of the channel.

The user, which is disposed at the receiving end, continuously senses the whole communication band. Considering the coexistence of the user signal and the jammer signal, the PSD of the signal at the receiving end can be expressed as:

$$R_{t}(f) = g^{TR} U\left(f - f_{t}^{I}\right) + g^{JR} J\left(f - f_{t}^{J}\right) + g^{ER} E(f) + n(f) \quad (3)$$

The discrete spectrum sample value is defined as $r_{n,t} = 10\log\left[\int_{n\Delta f}^{(n+1)\Delta f} R_{t}(f + f_{s}) df\right]$, where $\Delta f$ is the resolution of the spectrum analysis. The user determines the transmission frequency based on the spectrum vector $s_{t} = \{r_{1,t}, r_{2,t}, \cdots, r_{N,t}\}$.

## 3. An Anti-Jamming Deep Reinforcement Learning Algorithm based on Hiding strategy

Although the user does not know whether its actions avoid being sensed by the jammer, some patterns can be found in the actions of the jammer. The action-making process for jammer is shown in Fig. 2.

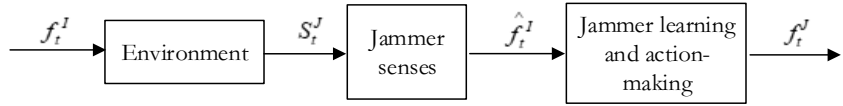

**Fig. 2.** The jammer's action-making process

From Fig. 2, it can be seen that at the moment $t$ when the user frequency action $f_{t}^{I}$ is entered into the environment, the jammer will sense the environmental state $S_{t}^{J}$, and identify the user's frequency $\hat{f}_{t}^{I}$. The jammer then learns from $\hat{f}_{t}^{I}$ and decides for the corresponding frequency $f_t^J$. Assuming that $\hat{f}_t^I = f_t^I$, the jammer's action sequence $\overrightarrow{Y_t} = \left( f_t^J, f_{t-1}^J, \cdots, f_{t-T+1}^J \right)$ must be highly correlated with that of the user's action sequence $\overrightarrow{X_t} = \left( f_t^I, f_{t-1}^I, \cdots, f_{t-T+1}^I \right)$. Conversely, it $\overrightarrow{Y_t}$ has little or no correlation with $\overrightarrow{X_t}$, it means that $\hat{f}_t^I \neq f_t^I$, i.e., the jammer does not correctly sense the user's action information.

From the above, it is clear that the user has no direct access to whether its decisions are avoid being sensed by the jammer. However, it is possible to obtain disturbed decision sequences by receiving environmental states. Comparing the correlation between the user's decision sequence $\overrightarrow{X_t}$ and the jammer's decision sequence $\overrightarrow{Y_t}$ indirectly verifies whether the user's decision avoid being sensed by the jammer. So, a vector difference value function is proposed to measure the correlation between the action sequences of the user and the jammer, whether the user's action avoids the jammer's sense can be judged. The vector difference value function $\rho_{\overrightarrow{X_t},\overrightarrow{Y_t}}(m)$ is shown below:

$$\rho_{\overrightarrow{X_t},\overrightarrow{Y_t}}(m) = \left| \overrightarrow{X_t} \bullet \overrightarrow{Y_{t+m}} \right|_d, m \in \left[ -(T-1), (T+1) \right] \quad (4)$$

where the vectors $\overrightarrow{X_t}$ and $\overrightarrow{Y_t}$ represent the action sequences of the user and the jammer, respectively, and $\left| \overrightarrow{X_t} \bullet \overrightarrow{Y_{t+m}} \right|_d$ is specifically calculated as follows:

$$\left| \overrightarrow{X_t} \bullet \overrightarrow{Y_{t+m}} \right|_d = \frac{T - \sum_{n=0}^{T-1} \delta_n}{T} \quad (5)$$

Where $\delta_n$ is defined by equation 6:

$$\delta_n = \begin{cases} 0, \left( f_{t-n}^I - f_{t-n+m}^J \right) - K_{\overrightarrow{X_t},\overrightarrow{Y_t}} = 0 \\ 1, \left( f_{t-n}^I - f_{t-n+m}^J \right) - K_{\overrightarrow{X_t},\overrightarrow{Y_t}} \neq 0 \end{cases} \quad (6)$$

Assume that $Z_{t-n} = f_{t-n}^I - f_{t-n+m}^J, n \in [0, T-1]$, where $K_{\overrightarrow{X_t},\overrightarrow{Y_t}}$ denotes the element in $\{Z_{t-n}\}$ with the highest probability of occurrence, i.e., the distance bias between the vectors $\overrightarrow{X_t}$ and $\overrightarrow{Y_t}$. It follows that the smaller the vector difference value function $\rho_{\overrightarrow{X_t},\overrightarrow{Y_t}}(m)$, the less correlation between the two action sequences and the better the hidden anti-jamming effect.

To make the vector difference value function $\rho_{\overrightarrow{X_t Y_t}}(m)$ achieve the best effect $R = max\left(\rho_{\overrightarrow{X_t Y_t}}(0), \rho_{\overrightarrow{X_t Y_t}}(1), \cdots, \rho_{\overrightarrow{X_t Y_t}}(m)\right)$ is defined to select the largest vector difference value function for the judgment.

As the environment is unknown and dynamic, it is impossible to obtain the available frequencies directly from the environment $s_t$. For intelligent jamming with complex interference pattern, the jammer's actions may be related to the previous historical state, so we define the environment state as $S_t = \{s_t, s_{t-1}, \cdots, s_{t-T+1}\}$, where $T$ denotes the length of historical states of backtracking [21]. In our anti-jamming Markov action process (MDP), $S_t \in \{S_1, S_2, S_3, \cdots\}$ is the temporal spread environment state, $a_t \in \{f_1^I, f_2^I, \cdots, f_N^I\}$ is the frequency action of the user, $P(S_{t+1} | S_t, a_t)$ is the transition probability from the current state $S_t$ to $S_{t+1}$ when taking action $a_t$.

The user needs to ensure the communication quality, therefore, this paper chooses the communication quality assessment based on the user's SINR ratio. Besides, the user wants its action information not to be sensed by the jammer, that is, the correlation between the user's action sequence and the jammer's action sequence is minimal, $R$ is the minimum. Combining these two factors, the instantaneous reward $r_t$ is defined by Equation 7, after choosing an action $a_t$ in the state $S_t$:

$$r_t = \log(1 + \Phi_t^I) + \alpha(1 - R) \tag{7}$$

where $\alpha$ denotes the correlation coefficient of action.

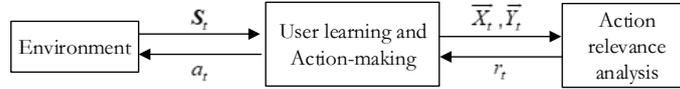

**Fig. 3.** Flowchart of action making on the user

The correlation analysis when the user makes an action is shown in Fig. 3, where the user receives the environmental state $S_t$ and makes action $a_t$ by learning. Simultaneously correlates the user's action sequence $\overrightarrow{X_t} = (f_t^I, f_{t-1}^I, \cdots, f_{t-T+1}^I)$ and the jammer's action sequence $\vec{Y_t} = (f_t^J, f_{t-1}^J, \cdots, f_{t-T+1}^J)$ in the environmental state $S_t$, which results in an instantaneous reward $r_t$. Adrift of the environmental state to the next moment is $S_{t+1}$.

During the iterative learning process, the user continuously interacts with the environment to explore the changing patterns of the jamming actions to obtain the optimal transmission actions. Considering that the expansion of the time dimension causes the state of the environment to become very large. The reinforcement learning converges too slowly and takes too long to apply directly. In this paper, the Q function is estimated by using a convolutional neural network (CNN). And the CNN has been verified in [24] that can effectively extract time-frequency features of jamming. Q function is the expected discounted long-term reward for state S and action a, i.e.,

$$Q(S_t, a_t) = E\left\{r_t + \gamma \max_{a_{t+1}} Q(S_{t+1}, a_{t+1}) | S_t, a_t\right\} \tag{8}$$

where $S_{t+1}$ denotes the next state if the user chooses action $a_t$ at the state $S_t$, $\gamma$ is the reward discount factor.

---

**Algorithm 1:** An anti-jamming deep reinforcement learning algorithm based on hiding strategy

**Initialize** $D=\emptyset$, $i=0$, network weight $\theta_0$ with random values, $Training = $ **True**, $S_1 = O(T \times N)$.
**For** $t = 1, 2, \cdots, \infty$ **do**
  **If** *Training* **then**
    Select action $a_t$ **via** $\varepsilon - \text{greedy}$ algorithm **or** Select action $a_t = \arg\max_a Q(S_t, a; \theta)$

    Execute $a_t$, record $r_t$ and sense $S_{t+1}$, store $e_t = (S_t, a_t, r_t, S_{t+1})$, Store $e_t$ in $D$.
    **If** $Sizeof(D) > N/2$
      Select $e_t$ from $D$ randomly,
      Compute $\eta_i = r_t + \gamma \max_{a_{t+1}} Q(S_{t+1}, a_{t+1}; \theta_{i-1})$,

      Compute $L_i(\theta_i) = E(\eta_i - Q(S_t, a_t; \theta_i))^2$,

      Compute $\nabla_{\theta_i} L_i(\theta_i)$, update $\theta_i$ with gradient descent algorithm, **and** $i = i+1$
    **End If**
  **End If**
  **If** $L_i(\theta_i) < \Gamma$
    $Training = $ **False**
    Select action $a_t = \arg\max_a Q(S_t, a; \theta')$

  **End If**
**End For**

Similar to the method in [24], environmental state $S_t$, action $a_t$, reward $r_t$, and environmental state $S_{t+1}$ are made to be the experience $e_t = (S_t, a_t, r_t, S_{t+1})$ obtained at the moment $t$. The experience of each moment is recorded in the dataset $D = (e_1, e_2, \cdots, e_t)$ until the dataset $D$ reaches a certain number. With equal probability, a set of experiences is extracted from the data set $D$ to construct the target value $\eta_i = r_t + \gamma \max_{a_{t+1}} Q(S_{t+1}, a_{t+1}; \theta_{i-1})$, where $\theta_i$ is the network weighting factor at the number iteration. The loss function $L_i(\theta_i) = E(\eta_i - Q(S_t, a_t; \theta_i))^2$ is calculated between the target value $\eta_i$ and the true value $Q(S_t, a_t; \theta_i)$. The gradient of the loss function $L_i(\theta_i)$ for the network weight factor is expressed as:

$$\nabla_{\theta_i} L_i(\theta_i) = E\left[L_i(\theta_i) \nabla_{\theta_i} Q(S_t, a_t; \theta_i)\right] \tag{9}$$

Ultimately, the network weights $\theta_i$ can be updated using stochastic gradient descent until the end of the algorithm. The ADRLH is shown in Algorithm 1.

At the beginning of the iteration, the algorithm is trained and the network weighting factor $\theta_i$ is updated to $\theta_{i+1}$. If $L(\theta_{i+1})$ is less than the threshold $\Gamma$, $\theta'\ (\theta' = \theta_{i+1})$ will no longer update and the algorithm select the optimal action $a_t = \arg\max_a Q(S_t, a; \theta')$.

## 4. Simulation Results and Analysis

### 4.1 Simulation parameter

In this paper, the user and the jammer combat with each other in a frequency band of 10MHz. And the number of channels $N=10$. The antagonists perform a spectral probe once per $1ms$, and their center frequency can be changed once per $10ms$ depending on the action. Both signal and jamming are raised cosine waveforms with roll-off factor α = 0.5, in which jamming power is 40dBm and signal power is 30dBm. The spectral resolution $\Delta f$ is set to be $1kHz$.

The user retraces the time duration $T = 100ms$. The environmental state $S_t$ is a two-dimensional matrix $10 \times 100$. The action correlation coefficient $\alpha = 2$. The reward discount factor is $\gamma = 0.8$. The results of 10,000 iterations were used, and the average of every 100 iterations was used as a data point.

Two kinds of jamming scenarios are considered for simulation, follower jamming and based on Q-learning jamming, respectively. ADRLH are compared with three methods, It contains Adaptive Frequency Hopping(AFH), Frequency Hopping Spread Spectrum(FHSS), and Anti-jamming Deep Reinforcement Learning Algorithm based on Avoiding strategy(ADRLA).

## 4.2 Simulation results and analysis

For illustration and presentation, we first give the receiving spectrum of the jammer and the user when the user adopts ADRLA and ADRLH. The sense spectrum of them when the user adopts ADRLA is shown in Fig. 4. The $(a)$ is the spectrum receiving by the jammer and $(b)$ is the spectrum receiving by the user. The result shows that when the user adopts ADRLA, the communication channel is selected by only considered the quality of communication. Although the user's actions can avoid jamming, the information of the user may be obtained by the jammer.

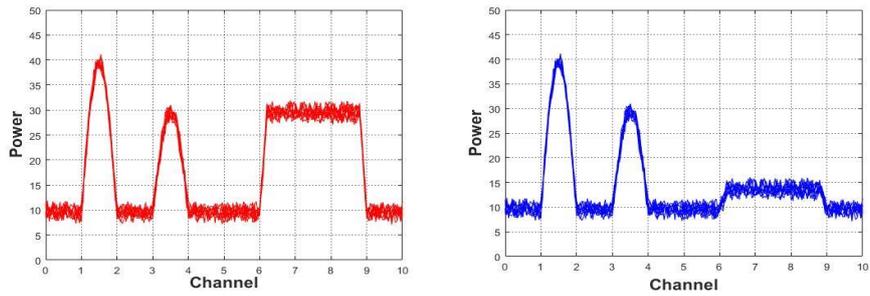

(a) receiving spectrum of the jammer      (b) receiving spectrum of the user

**Fig. 4**. Receiving spectrum of the jammer and the user based on ADRLA

The receiving spectrum of them when the user adopts ADRLH is shown in Fig. 5. It is clear from the comparison between $(a)$ and $(b)$ that the actions is made not only on the quality of communication but also on the selection of channel that is not sensed by the jammer. Although the quality of communication will be slightly affected, it avoids the user's information leakage.

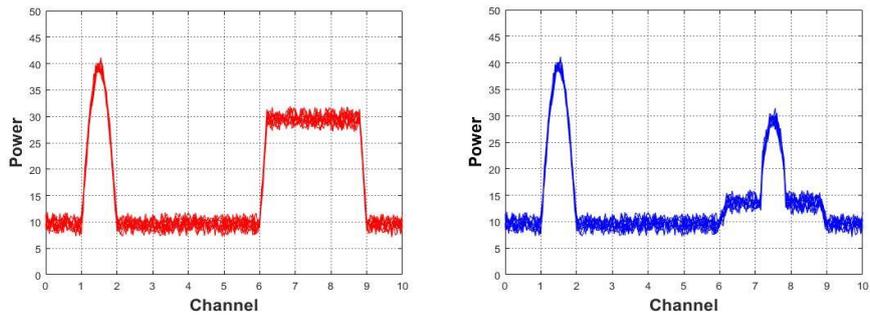

(a) receiving spectrum of the jammer      (b) receiving spectrum of the user

**Fig. 5**. Receiving spectrum of the jammer and the user based on ADRLH

Secondly, the sensing probability of jammer with the iterations is shown in Fig.6. When the user adopts AFH method, the probability that being sensed by jammer is higher than that of the other three methods. At the beginning of the iteration, there is no obvious difference between the probability being sensed by jammer when user adopts ADRLH, ADRLA or FHSS. As the number of iterations increases, the sensing probability of jammer decreases if the user adopts ADRLH but increases when adopting ADRLA. Compared with the other three methods, the proposed algorithm reduces the sensing probability of jammer by 82.7%, 59.2%, and 75.8%, respectively. Fig.7. depicts the relationship between the action correlation coefficient and the number of iterations. Comparing Fig.6. with Fig.7., it can be seen that no matter which method the user adopts, as the number of iterations increases, the changing trends of the correlation between the actions of the user and the jammer and the sensing probability of jammer are consistent. In summary, the action correlation measurement method can verify whether the user's actions avoid being sensed by the jammer.

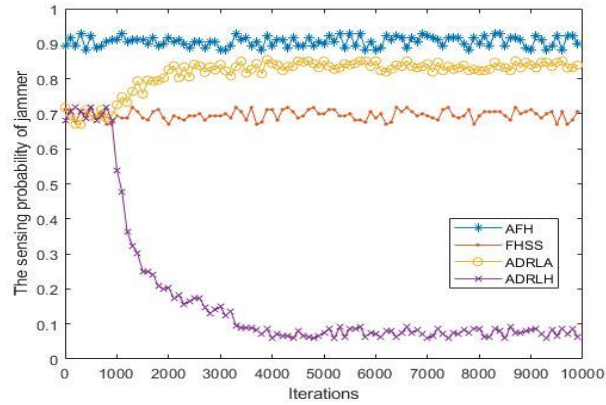

**Fig. 6.** The sensing probability of jammer

Thirdly, the normalized throughput of the system with the iterations is shown in Fig. 8. At it beginning of learning procedure, the normalized throughput performance of ADRLH, FHSS, and ADRLA are close to each other, and are superior to AFH. As the iteration progresses, the normalized throughput of ADRLH and ADRLA will gradually increase while AFH will remain unchanged and FHSS will gradually decrease. When steady converging, compared with other methods, the normalized throughput of ADRLH is increased by 0.124, 0.332, and 0.687 respectively.

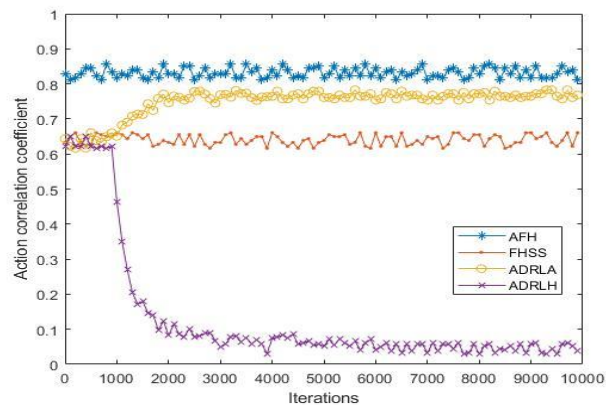

**Fig.7**. Relationship between the action correlation coefficient and iterations

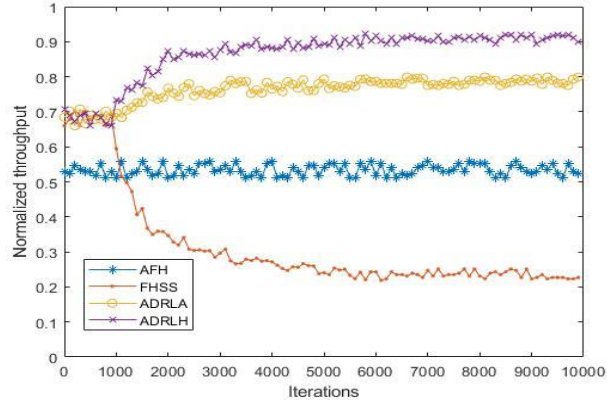

**Fig.8**. Normalized throughput of the system with iterations

Finally, Table 4-1 shows the probability of the user being jammed when different anti-jamming methods deal with follower jamming and Q-learning-based jamming. It is seen from Table 4-1 that performance of the hidden anti-jamming method proposed in this paper is better than that of AFH and FHSS when dealing with follower jamming or intelligent jamming based on Q-learning, i.e. ADRLH reduced the probability being jammed by 14.7% compared to ADRLA when fighting against intelligent jamming based on Q-learning.

**Table 4-1** Probability of jamming

| Jamming methods<br>Anti-Jamming methods | Follower jamming | Q-Learning |
|:---:|:---:|:---:|
| AFH | 45.7648% | 25.6359% |
| FHSS | 8.6523% | 76.8216% |
| ADRLA | 1.0867% | 23.4368% |
| ADRLH | 0.9489% | 8.7475% |

# 5. Conclusion

This paper proposes a hidden anti-jamming method based on the idea of reducing the sensing probability of the jammer. A deep reinforcement learning framework is designed. By analyzing the correlation between the actions of the user and the jammer, the sensing probability of the jammer is indirectly obtained. To measure the action correlation between the user and the jammer, an action correlation measurement method is designed. Combined with the deep reinforcement learning method, Anti-jamming Deep Reinforcement Learning Algorithm based on Hiding strategy(ADRLH) is proposed, aiming to avoid the information leakage of the user under the premise of ensuring communication quality. Simulation results show that AD RLH is improved in jamming avoidance and much improved to avoid being sensed by the jammer compared with the algorithms that only consider jamming avoidance.

# References

[1] A. Kavianpour and M. C. Anderson, "An Overview of Wireless Network Security," 2017 IEEE 4th International Conference on Cyber Security and Cloud Computing (CSCloud), New York, NY, 2017, pp. 306-309, doi: 10.1109/CSCloud.2017.45.

[2] J. Li, Z. Feng, Z. Feng and P. Zhang, "A survey of security issues in Cognitive Radio Networks," in China Communications, vol. 12, no. 3, pp. 132-150, Mar. 2015, doi: 10.1109/CC.2015.7084371.

[3] M. Bkassiny, Y. Li and S. K. Jayaweera, "A Survey on Machine-Learning Techniques in Cognitive Radios," in IEEE Communications Surveys & Tutorials, vol. 15, no. 3, pp. 1136-1159, Third Quarter 2013, doi: 10.1109/SURV.2012.100412.00017.

[4] X. Wang et al., "Dynamic Spectrum Anti-Jamming Communications: Challenges and Opportunities," in IEEE Communications Magazine, vol. 58, no. 2, pp. 79-85, February 2020, doi: 10.1109/MCOM.001.1900530.

[5] O. Naparstek and K. Cohen, "Deep Multi-User Reinforcement Learning for Distributed Dynamic Spectrum Access," in IEEE Transactions on Wireless Communications, vol. 18, no. 1, pp. 310-323, Jan. 2019, doi: 10.1109/TWC.2018.2879433.

[6] H. Zhu, C. Fang, Y. Liu, C. Chen, M. Li and X. S. Shen, "You Can Jam But You Cannot Hide: Defending Against Jamming Attacks for Geo-Location Database Driven Spectrum Sharing," in IEEE Journal on Selected Areas in Communications, vol. 34, no. 10, pp. 2723-2737, Oct. 2016, doi: 10.1109/JSAC.2016.2605799.

[7] J. Xu and N. Wang, "Optimization of ROV Control Based on Genetic Algorithm," 2018 OCEANS - MTS/IEEE Kobe Techno-Oceans (OTO), Kobe, 2018, pp. 1-4, doi: 10.1109/OCEANSKOBE.2018.8559384.

[8] Eryong Yang, Jianzhong Chen and Yingtao Niu, "Anti-jamming communication action engine based on Particle Swarm Optimization," 2011 Second International Conference on Mechanic Automation and Control Engineering, Hohhot, 2011, pp. 3913-3916, doi: 10.1109/MACE.2011.5987855.

[9] Xianyang Hui, Yingtao Niu and Mi Yang, "Decision method of anti-jamming communication based on binary artificial bee colony algorithm," 2015 4th International Conference on Computer Science and Network Technology (ICCSNT), Harbin, 2015, pp. 987-991, doi: 10.1109/ICCSNT.2015.7490902.

[10] B. Wang, Y. Wu, K. J. R. Liu and T. C. Clancy, "An anti-jamming stochastic game for cognitive radio networks," in IEEE Journal on Selected Areas in Communications, vol. 29, no. 4, pp. 877-889, April 2011, doi: 10.1109/JSAC.2011.110418.

[11] L. Jia, Y. Xu, Y. Sun, S. Feng, L. Yu and A. Anpalagan, "A Multi-Domain Anti-Jamming Defense Scheme in Heterogeneous Wireless Networks," in IEEE Access, vol. 6, pp. 40177-40188, 2018, doi: 10.1109/ACCESS.2018.2850879.

[12] Y. Xu, J. Wang, Q. Wu, J. Zheng, L. Shen and A. Anpalagan, "Dynamic Spectrum Access in Time-Varying Environment: Distributed Learning Beyond Expectation Optimization," in IEEE Transactions on Communications, vol. 65, no. 12, pp. 5305-5318, Dec. 2017, doi: 10.1109/TCOMM.2017.2734768.

[13] Y. Xu, J. Wang, Q. Wu, A. Anpalagan and Y. Yao, "Opportunistic Spectrum Access in Unknown Dynamic Environment: A Game-Theoretic Stochastic Learning Solution," in IEEE Transactions on Wireless Communications, vol. 11, no. 4, pp. 1380-1391, April 2012, doi:

10.1109/TWC.2012.020812.110025.

[14] L. Jia, Y. Xu, Y. Sun, S. Feng and A. Anpalagan, "Stackelberg Game Approaches for Anti-Jamming Defence in Wireless Networks," in IEEE Wireless Communications, vol. 25, no. 6, pp. 120-128, December 2018, doi: 10.1109/MWC.2017.1700363.

[15] J. Parras, J. del Val, S. Zazo, J. Zazo and S. V. Macua, "A new approach for solving anti-jamming games in stochastic scenarios as pursuit-evasion games," 2016 IEEE Statistical Signal Processing Workshop (SSP), Palma de Mallorca, 2016, pp. 1-5, doi: 10.1109/SSP.2016.7551804.

[16] Z. Feng et al., "Power Control in Relay-Assisted Anti-Jamming Systems: A Bayesian Three-Layer Stackelberg Game Approach," in IEEE Access, vol. 7, pp. 14623-14636, 2019, doi: 10.1109/ACCESS.2019.2893459.

[17] Y. Wang, Y. Niu, J. Chen, F. Fang and C. Han, "Q-Learning Based Adaptive Frequency Hopping Strategy Under Probabilistic Jamming," 2019 11th International Conference on Wireless Communications and Signal Processing (WCSP), Xi'an, China, 2019, pp. 1-7, doi: 10.1109/WCSP.2019.8927884.

[18] X. Pei et al., "Joint Time-frequency Anti-jamming Communications: A Reinforcement Learning Approach," 2019 11th International Conference on Wireless Communications and Signal Processing (WCSP), Xi'an, China, 2019, pp. 1-6, doi: 10.1109/WCSP.2019.8928061.

[19] Z. Zhang, Q. Wu, B. Zhang and J. Peng, "Intelligent Anti-Jamming Relay Communication System Based on Reinforcement Learning," 2019 2nd International Conference on Communication Engineering and Technology (ICCET), Nagoya, Japan, 2019, pp. 52-56, doi: 10.1109/ICCET.2019.8726916.

[20] S. Liu et al., "Pattern-Aware Intelligent Anti-Jamming Communication: A Sequential Deep Reinforcement Learning Approach," in IEEE Access, vol. 7, pp. 169204-169216, 2019, doi: 10.1109/ACCESS.2019.2954531.

[21] G. Han, L. Xiao and H. V. Poor, "Two-dimensional anti-jamming communication based on deep reinforcement learning," 2017 IEEE International Conference on Acoustics, Speech and Signal Processing (ICASSP), New Orleans, LA, 2017, pp. 2087-2091, doi: 10.1109/ICASSP.2017.7952524.

[22] X. Liu, Y. Xu, Y. Cheng, Y. Li, L. Zhao and X. Zhang, "A heterogeneous information fusion deep reinforcement learning for intelligent frequency selection of HF communication," in China Communications, vol. 15, no. 9, pp. 73-84, Sept. 2018, doi: 10.1109/CC.2018.8456453.

[23] X. Liu, Y. Xu, L. Jia, Q. Wu and A. Anpalagan, "Anti-Jamming Communications Using Spectrum Waterfall: A Deep Reinforcement Learning Approach," in IEEE Communications Letters, vol. 22, no. 5, pp. 998-1001, May 2018, doi: 10.1109/LCOMM.2018.2815018.

[24] W. Li, J. Wang, L. Li, G. Zhang, Z. Dang and S. Li, "Intelligent Anti-Jamming Communication with Continuous Action Decision for Ultra-Dense Network," ICC 2019 - 2019 IEEE International Conference on Communications (ICC), Shanghai, China, 2019, pp. 1-7, doi: 10.1109/ICC.2019.8761578.

[25] Y. Li, Y. Xu, X. Wang, W. Li and W. Bai, "Power and Frequency Selection optimization in Anti-Jamming Communication: A Deep Reinforcement Learning Approach," 2019 IEEE 5th International Conference on Computer and Communications (ICCC), Chengdu, China, 2019, pp. 815-820, doi: 10.1109/ICCC47050.2019.9064174.